\documentclass[10pt]{iopart}
\usepackage{epsf}
\usepackage{fleqn}

\newcommand{\nn}{\nonumber}
\newcommand{\be}{\begin{equation}}
\newcommand{\ee}{\end{equation}}
\newcommand{\bea}{\begin{eqnarray}}
\newcommand{\eea}{\end{eqnarray}}

\newcommand{\XI}{\mbox{\boldmath $\xi$}}
\newcommand{\x}{\mbox{\boldmath $x$}}
\newcommand{\w}{\mbox{\boldmath $w$}}

\newcommand{\gammabar}{\bar{\gamma}}
\newcommand{\gammabd}{\hat{\bar{\gamma}}_i^{\mu}(b,\w_i)}

\begin{document}

\title{Probing the basins of attraction of a recurrent neural network}
\author{M Heerema \footnote[6]{E-mail address: heerema@wins.uva.nl} and WA
    van Leeuwen \footnote[7]{E-mail address: leeuwen@wins.uva.nl} }
\address{Institute for Theoretical Physics, University of Amsterdam,
  Valckenierstraat 65, 1018 XE Amsterdam, The Netherlands}

\abstract
A recurrent neural network is considered that can retrieve a
collection of patterns, as well as slightly perturbed versions of
this `pure' set of patterns via fixed points of its dynamics.
By replacing the set of dynamical constraints, i.e., the fixed point
equations, by an extended collection of fixed-point-like equations, analytical expressions
are found for the weights $w_{ij}(b)$ of the net, which depend on a
certain parameter $b$.
This so-called basin parameter $b$ is such that for $b=0$ there are, a
priori, no
perturbed patterns to be recognized by the net.
It is shown by a numerical study, via probing sets, that a net
constructed to recognize perturbed patterns, i.e., with values of the
connections $w_{ij}(b)$ with $b \neq 0$, possesses larger basins of
attraction than a net made with the help of a pure set of patterns,
i.e., with connections $w_{ij}(b=0)$.
The mathematical results obtained can, in principle, be realized by an
actual, biological neural net.

\endabstract


\section{Introduction}
\label{introduction-basins}

The capacity of a neural network to recognize a pattern that is not
precisely equal to, but resembles, a given, stored pattern, is
characterized by what is called, in a mathematical context, the `basin
of attraction' of the stored pattern.
If the basin is small, the network will be capable only to associate a
small set of similar patterns to a typical pattern, whereas for a
large basin the set of similar patterns that can be recognized is
large.

Once a pattern has been presented to a neural network, the neural
network starts to evolve under the influence of its own internal dynamics.
If the network, at the end of this process, ends in a unique state
this state is called a (single) attractor of the network.
It is also possible that the network hops between more than one final state,
in which case one speaks of a multiple attractor \cite{bastolla,wongho,wong}.
Patterns that evolve to an attractor are said to belong to the basin
of attraction of this attractor.
Many ways of characterizing basins are en vogue: basins are
said to be deep or shallow and narrow or wide \cite{ws}. 

A way to influence the basins of the attractors is to change the network dynamics,
switching from deterministic to stochastic dynamics \cite{wong-sher2,AEHW}.
Another way to change the dynamics of the neural network is to vary
the connections during the learning stage.
The latter possibility can be exploited in a model for an actual, biological
system \cite{wong-sher3,wong-sher,gsw}.

We are primarily interested in biological neural networks.
Therefore, we are not aiming at mathematical problems such as
(optimizing) the storage capacity in relation to the sizes of the basins
of attraction, a subject that has got ample attention in the
literature \cite{kitano,neto+fon2,erichsen,yw,g89}.

Many dynamical systems are parameterized by a certain constant $\kappa$, sometimes
called `margin parameter'.
The margin parameter $\kappa$ is claimed to be related to the size of
the basins of attraction of the fixed points of the dynamics of a
neural network \cite{opper,krauth,gardner,kepler,forrest}.
Naively, one would expect, for reasons that are directly related to
the way this parameter $\kappa$ is introduced in the model that, the
larger $\kappa$, the larger the basins of attraction will be.
However, as the $1997$ study of Rodrigues Neto and Fontanari indicates
this may not be true.
Their numerical analysis, for tiny networks (up to $24$ neurons)
suggest that the number of attractors increase with increasing
$\kappa$ and that, perhaps because of this increase, the basins of
attraction are not enlarged --- as one might expect in first
instance \cite{neto+fon}.
In section~\ref{solving}, we arrive at a precise interpretation for the
margin parameter $\kappa$, which, usually, is introduced as an ad hoc
quantity.
In section~\ref{num-results}, we consider the effect of the margin parameter on the basins for a
network with $256$ neurons.
For a basin parameter $b=0$, we find that the larger $\kappa$, the
larger the basins, in agreement with what one would expect naively
(see figure~\ref{fignoise1} for $b=0$).

The $1992$ study of Wong and Sherrington is also concerned with the
sizes of the basins of attraction. One of their finding is, roughly
speaking, that the noisier the set of learning patterns, the larger
the basins of attraction \cite{ws}.
Our findings support their observations (see figure~\ref{fignoise1}
for $b>0$).

We consider a network in its final state only, i.e., after the process
of learning has stopped.
This makes our study time-independent.
We try and construct a network with weights $w_{ij}$ that can not only
store a certain set of $p$ prescribed patterns
$\XI^{\mu}=(\xi_1^{\mu},\ldots,\xi_N^{\mu})$, where $\mu=1,\ldots,p$,
but that can also remember a larger set of patterns, centered around
these typical patterns.
These enlarged sets, called $\Omega^{\mu}(b)$ below, are characterized
by the basin parameter $b$ mentioned above.
If $b=0$, the set $\Omega^{\mu}(b)$ reduces to the sole pattern
$\XI^{\mu}$.
What we obtain, finally, are values for the weights that depend on this basin
parameter $b$:
\be
\label{main}
w_{ij}(t) =  \cases{w_{ij}(t_0)+ v_{ij}(t) & ($j \in V_i$)\\ 
w_{ij}(t_0) & ($j \in V_i^c$) }
\ee
with
\be
\fl
\label{v-main}
v_{ij}(t)= N^{-1} \sum_{\mu,\nu=1}^p [\kappa -
  \bar{\gamma}_i^{\mu}(b,\w_i(t_0))](2 \xi_i^{\mu}-1)  (\bar{C}_i^{-1}(b))^{\mu
  \nu} [ (1-b) \xi_j^{\nu} + b(1-\xi_j^{\nu})]
\ee
[see section~\ref{otherwork}, eqs. (\ref{w-learn}), (\ref{v-alt}) and
  (\ref{v-final})].
Here, the $w_{ij}(t_0)$ are arbitrary numbers, which can be interpreted,
  in a different context, as initial values for the weights, at an
  initial time $t_0$, as is suggested by the notation.
Furthermore, $N$ is the number of neurons of the network.
We abbreviated $\w_i(t_0):=(w_{i1}(t_0),\ldots,w_{iN}(t_0))$. 
The quantities
  $\bar{\gamma}_i^{\mu}$ are defined in (\ref{gamma-basins}) for
  arbitrary $\w_i$, and the matrices
  $\bar{C}_i^{\mu \nu}$ are defined in (\ref{c}).
The $\bar{\gamma}_i^{\mu}$ depend on threshold potentials $\theta_i$,
  the basin parameter $b$ and the input patterns $\XI^{\mu}$.
$V_i$ and $V_i^c$ are index collections defined in such a way that the
  weights $w_{ij}$ are adaptable if $j \in V_i$ and constant if $j \in
  V_i^c$ (for all $i=1,\ldots,N$).
If the (constants) $w_{ij}$ vanish, there is no connection between
  $i$ and $j$.
Hence, the $w_{ij}$ ($j \in V_i^c$) determine the network topology.
The more $w_{ij}$ ($j \in V_i^c$) vanish, the more `diluted' a network.
We have written $w_{ij}(t)$ for the weights, to
  facilitate comparison with earlier result (see, e.g.,
  \cite{heerema}).
In the present article they are time independent constants, however.

The formulae (\ref{main})--(\ref{v-main}), constituting the main result of this article, generalize
well-known results for the weights of a recurrent network.
The generalizations concerned are: the network may not be fully connected,
and the weights may depend upon the prescribed sizes of the basins,
characterized by the basin parameter $b$.
For $b=0$ we recover our earlier result for a diluted network~\cite{heerema}.

It turns out that in some cases the basins of attraction are larger
for values of $b$ unequal to zero.
In other words, a network which has learned not only a set of patterns
$\XI^{\mu}$, but a collection of perturbed patterns $\Omega^{\mu}(b)$,
will possess larger basins of attraction.
Hence, a network can optimally recognize perturbed
patterns, if it has been constructed with perturbed patterns.
This is what Wong and Sherrington \cite{ws}, in a related study, but
for a network with connections that are changed during a learning
process, call the `principle of adaptation': a neural network is found to
perform best in an operating environment identical to the training
environment.
Our analysis of the system after the process of learning is completed
confirms this observation, albeit that the word `identical' is not to
be taken literally. 
So far our general introduction to the problem.
We now come to a short overview of our article.
   
In section~\ref{math-form} we start by defining mathematically the
problem to find suitable synaptic weights by
formulating the equations to be obeyed by the weights $w_{ij}$ of the
connections. 
In section~\ref{right} and the appendix we indicate how we could
obtain, in principle, a series expansion in the parameter $b$ for the
solution of the equations.
To actually calculate the first terms of the expansion would be very
time consuming.
We therefore proceed differently.
In section~\ref{alt-impl} we rewrite the implicit expression found in
section~\ref{right} in such a way, that we can easily find an
approximation [see eq.~(\ref{w-expl})]. 
What we essentially do, is to replace in the alternative implicit
expression found in section~\ref{alt-impl}, a certain average
$\bar{\gamma}_i^{\mu}$ related to the
$i$-th neuron potential $h_i$, threshold $\theta_i$ and the activity
$x_i$, given explicitly by eq.~(\ref{def-gamma}) below, by one and the same
constant $\kappa$. 
We thus find, by identification of $\kappa$ in an old result and the
$\kappa$ introduced here, an interpretation for the margin parameter $\kappa$.
Whether this replacement of the functions $\bar{\gamma}_i^{\mu}$ by one and
the same constant makes sense, is studied in the next section.
In section~\ref{num-results} we introduce a probing set, characterized
by a probing parameter $\bar{b}$.
The network's performance, as a function of the basin parameter $b$, is
calculated numerically for different values of the probing parameter
$\bar{b}$. 
We thus test our approximation to the exact solution, and find it to
be quite satisfactory.

\section{Mathematical formulation of the problem}
\label{math-form}

\subsection{Equations for the enlarged sets of input patterns}
\label{eq-basin}

Consider a recurrent network of $N$ neurons.
A neuron $i$ of this network fires ($x_i=1$) if its potential
$h_i=\sum_{l=1}^N w_{il} x_l$ surpasses a
certain threshold value $\theta_i$ ($i=1,\ldots,N$).
The dynamics of the network is given by the deterministic equation
\be
\label{dyn-rule-basins}
x_i(t+\Delta t)= \Theta_{\rm H}(\sum_{l=1}^N w_{il}x_l(t)-\theta_i) \qquad (i=1,\ldots,N ) 
\ee
where $\Theta_{\rm H}$ is the Heaviside step function: $\Theta_{\rm H}(z)=1$ for
  positive $z$ and zero elsewhere.
The weights of the neurons of the network will be updated
simultaneously, i.e., we use parallel dynamics.

Let us suppose that the network is such that it can store the $p$
patterns $\XI^{1},\ldots,\XI^{p}$, where $\XI$ is an $N$ dimensional
vector consisting of zeros and ones.
For the $\mu$-th pattern we have $h_i=\sum_{l=1}^N w_{il}\xi_l^{\mu}$, hence
the weights $w_{il}$ of this network are constrained by the fixed
point equations, following from (\ref{dyn-rule-basins})
\be
\label{fixed}
\Theta_{\rm H}(\sum_{l=1}^N w_{il}\xi_l^{\mu}-\theta_i)= \xi_i^{\mu}   \qquad (\mu=1,\ldots,p \, ; i=1,\ldots,N )
\ee
(see, e.g., \cite{heerema}).

Once the weights $w_{il}$ occurring in (\ref{fixed}) have been
determined for chosen collections of patterns $\XI^{\mu}$, one may ask
the question which patterns $\x$, alike but not exactly
equal to one of the $\XI^{\mu}$'s, evolve to the fixed point
$\XI^{\mu}$, i.e., what are the basins of attraction of the fixed
points $\XI^{\mu}$.
It is precisely the purpose of this article to study this question in
some detail.

The basin of attraction of an attractor of a dynamical system is defined to be the
collection of vectors that evolve, in one or many steps, to this attractor.
We are here interested in the question which vectors $\x$, belonging to
certain disjunct sets of patterns $\Omega^{\mu}$, centered around typical
patterns $\XI^{\mu}$ ($\mu=1,\ldots,p$), arrive, in one step of the
dynamics only, at the fixed point $\XI^{\mu}$.
These latter $\x$'s belong certainly to the basin of attraction as defined
above, and will be referred to, for the sake of simplicity, as the
basin of attraction, although it is only a part, namely, the one step
part, of the actual basin of attraction.
 
In order to take our newly defined basin of attraction into account,
we shall replace the requirement (\ref{fixed}), an equation for
the weights $w_{il}$, by
\be
\label{noise-basins}
\Theta_{\rm H}(\sum_{l=1}^N w_{il}x_l-\theta_i)= \xi_i^{\mu}
\qquad (\mu=1,\ldots,p \, ; i=1,\ldots,N ) 
\ee
where the patterns $\x$ belong to certain given disjunct sets of patterns
$\Omega^{\mu}$, still to be specified, centered around typical
patterns $\XI^{\mu}$ ($\mu=1,\ldots,p$). 
Equation (\ref{noise-basins}) is the central equation of this article; we are no
longer concerned with the equations (\ref{dyn-rule-basins}) or
(\ref{fixed}).
Note that this equation is time-independent; nevertheless, we will
indicate the final solution for the weights by $w_{ij}(t)$, in order
to suggest that these are the weights after a period of learning.
We shall determine by an (approximating) {\em analytical} procedure the weights
$w_{il}$ such that (\ref{noise-basins}) is probably satisfied for most of
the patterns $\x$, but not necessarily for all patterns.
The latter will depend on the chosen collections $\Omega^{\mu}$
($\mu=1,\ldots,p$). 
Having obtained the weights $w_{il}$, for such a particular choice of
$\Omega^{\mu}$'s, we shall check by a {\em numerical} procedure, whether all $\x \in
\Omega^{\mu}$ actually satisfy (\ref{noise-basins}).
This will indeed not always, but often, be the case.
Thus we shall have obtained values for the weights which could be useful for
an actual network.

As stated above, we are not concerned, in this article, with the
process via which learning takes place, we are only studying the
purely mathematical problem of finding values for the weights $w_{ij}$
that guarantee storage and retrieval properties of a neural net.
This leaves us with the question whether the values, given by such a
dry, mathematical requirement can actually be realized by the wet-ware
constituted by the neurons and their connections.
This point will be the subject of a next article \cite{heerema3}, where it
will turn out that a biological system can realize values for the
weights which very closely approximate the values obtained here:
compare formulae (\ref{main})--(\ref{v-main}) with ($38$)--($39$) of
\cite{heerema3}. 
 
Distinguishing the cases $\xi_i^{\mu}=0$ (no neuron activity) and $\xi_i^{\mu}=1$ one may
verify that the equations (\ref{noise-basins}) are fulfilled if and only if 
\be
\label{noise-alt}
\gamma_i^{\mu}(\w_i):=(\sum_{l=1}^N w_{il}x_l-\theta_i)(2 \xi_i^{\mu}-1)>0
\qquad \forall \, \x \in \Omega^{\mu} 
\ee 
where we abbreviated $\w_i:=(w_{i1},\ldots,w_{iN})$, and where $\Omega^{\mu}$ is a collection of patterns which will be made
explicit in section~\ref{pat-distr}.
Let $p^{\mu}(\x)$ be the probability of occurrence of a
pattern $\x$ in the set $\Omega^{\mu}$ of patterns centered around a typical pattern
$\XI^{\mu}$.
From (\ref{noise-alt}) it follows that the averages $\gammabar_i^{\mu}$ defined as
\be
\label{def-gamma}
\gammabar_i^{\mu}(\w_i):=\sum_{\x \in \Omega^{\mu}} p^{\mu}(\x) (\sum_{l=1}^N
w_{il}x_l-\theta_i)(2 \xi_i^{\mu}-1)  \qquad (\mu=1,\ldots,p \, ;i=1,\ldots,N) 
\ee
are also positive, i.e., 
\be
\label{noise-alt2}
\gammabar_i^{\mu}(\w_i) >0 \, .
\ee
Conversely, the fact that the averages are positive,
$\gammabar_i^{\mu}>0$, does not necessarily imply that
$\gamma_i^{\mu}>0$ ($i=1,\ldots,N; \mu=1,\ldots,p$).
Throughout this article, the averages $\gammabar_i^{\mu}$ will
play a central role. 

Let $n_{\Omega}$ be the total number of patterns belonging to any of
the collections $\Omega^{\mu}$ ($\mu=1,\ldots,p$).
Since, in general, the number $n_{\Omega}$ is larger than the number $p$,
the set of equations (\ref{noise-basins}) will be more restrictive than the set
(\ref{fixed}).
 
In the following, we shall consider biological networks, for which
$w_{ii}=0$ ($i=1,\ldots,N$).
Moreover, we shall consider partially connected (or diluted) networks,
i.e., we allow for the possibility that a particular set of
connections $w_{ij}$ vanish.
In general, we shall suppose that a certain subset of connections
$w_{ij}$ have prescribed values, which may or may not be zero.
In order to formalize this, we introduce the sets $V_i$
($i=1,\ldots,N$) and their complements $V_i^c$: the $V_i$ contain all
indices $j$ for which $w_{ij}$ is not prescribed, but to be determined
via equation (\ref{noise-basins}), while their
complements $V_i^c$ contain all indices $j$ for which $w_{ij}$ have
certain prescribed values, which may or may not be zero \cite{heerema}.
Let the total number of indices $j$ for which $w_{ij}$
($i=1,\ldots,N$) is prescribed be given by $M$.
Then (\ref{noise-basins}) is a set of $N n_{\Omega}$ inequalities to
be satisfied by the $N^2-M$ unknown weights $w_{ij}$.

Multiplying both sides of (\ref{noise-basins}) by $p^{\mu}(\x) x_j$
and summing over $\mu$ and $\x$ we obtain 
\be
\label{averaging}
\sum_{\mu=1}^p  \sum_{ \x \in \Omega^{\mu}}
p^{\mu}(\x) x_j \Theta_{\rm H}( \sum_{l=1}^N w_{il} x_l-\theta_i) =
\sum_{\mu=1}^p  \sum_{\x \in \Omega^{\mu}}
p^{\mu}(\x) x_j \xi_i^{\mu}  \qquad (j \in V_i) \, .
\ee
These are $N^2-M$ equations for the $N^2-M$ non-prescribed weights
$w_{ij}$, from which we want to solve the $w_{ij}$, once the
$\Omega^{\mu}$, or, equivalently, the $p^{\mu}(\x)$ are specified.
Notwithstanding the fact that the number of equations equals the number
of unknowns, the solution of (\ref{averaging}) for the weights
$w_{il}$ is not unique, because the step function $\Theta_H$ only
requires that $\sum_{l=1}^N w_{il} x_l - \theta_i$ be positive or negative.
As a side-remark we notice that equation (\ref{fixed}) is
under-determined for $p<N$: then there are more unknowns $w_{il}$ than
equations. 

\subsection{The distribution of patterns in the basins}
\label{pat-distr}

We choose the following, particular, probability distribution function
\be
\label{prob-basins}
p^{\mu}(\x)=\prod_{i=1}^N p_i^{\mu}(x_i) 
\ee
where
\be
\label{prob-i-basins}
p_i^{\mu}(x_i)=(1-b) \, \delta_{x_i,\xi_i^{\mu}}+b \, \delta_{x_i,1-\xi_i^{\mu}}
\ee
and where $b$ is a parameter between $0$ and $1$, which we will refer
to as the `basin-parameter'.

The sets $\Omega^{\mu}$ around the patterns $\XI^{\mu}$ are supposed
to be disjunct, and a vector $\x$ outside $\cup_{\mu=1}^p
\Omega^{\mu}$ has, by definition, a vanishing probability of
occurrence.
The probability distribution
(\ref{prob-basins})--(\ref{prob-i-basins}), however, yields a finite
--- albeit it very small --- probability of occurrence for a vector
$\x$ outside the direct surrounding $\Omega^{\mu}$ of $\XI^{\mu}$,
since it is defined for all $2^{N}$ possible vectors $\x$.
The observation that the probability distribution
(\ref{prob-basins})--(\ref{prob-i-basins}) for $\x$'s outside $\Omega^{\mu}$
is very small allows us to approximate the sum of all $\x \in
\cup_{\mu=1}^p \Omega^{\mu}$ by the larger sum over all $\x \in
\{0,1\}^N$. 
This approximation will enable us to obtain analytical results.

If $b=0$, only the patterns $\x=\XI^{\mu}$ have a non-zero
probability of occurrence.
For values of $b$ close to zero any vector $\x$ has a non-zero
probability of occurrence, but only vectors $\x$ close to one of the
$\XI^{\mu}$ have a probability of occurrence comparable to the
probability of occurrence of a typical pattern.
Note, that the basin-parameter is directly related to the magnitude of
$\Omega^{\mu}$: the larger $b$, the larger the number of patterns in
$\Omega^{\mu}$ that resemble the pattern $\XI^{\mu}$. 
Let us denote the average over the patterns as
\be
\label{average-x}
\bar{x}_j^{\mu}:=\sum_{\x \in \{0,1\}^N} p^{\mu}(\x) x_j \, .
\ee
Then, from (\ref{prob-basins}) and (\ref{prob-i-basins}) we find 
\be
\label{distr-1}
\sum_{\x \in \{0,1\}^N} p^{\mu}(\x)= 1
\ee
and
\bea
\fl
\label{distr-x-basins}
\bar{x}_j^{\mu} &=& \sum_{x_j=0,1}
\left[ (1-b) \, \delta_{x_j,\xi_j^{\mu}}+b \,
  \delta_{x_j,1-\xi_j^{\mu}} \right] x_j  \prod_{k \neq j} \sum_{x_k=0,1} \left[ (1-b) \,
\delta_{x_k,\xi_k^{\mu}}+b \, \delta_{x_k,1-\xi_k^{\mu}} \right] \nn\\
&=& (1-b) \xi_j^{\mu} +b (1-\xi_j^{\mu}) \, .
\eea
The first equation, eq. (\ref{distr-1}), expresses the normalization of the probability
distribution function, the second one, eq. (\ref{distr-x-basins}), expresses the fact that the
average value of the activity of neuron $j$ is a number between $0$
and $1$, depending on the basin-parameter $b$.
Using (\ref{average-x}) and (\ref{distr-1}) in (\ref{def-gamma}) yields
\be
\label{gamma-basins}
\gammabar_i^{\mu}(b,\w_i)= (\sum_{l=1}^N w_{il} \bar{x}_l^{\mu} -\theta_i)(2 \xi_i^{\mu}-1) 
\ee
where $b$ is the basin-parameter and where $\bar{x}_l^{\mu}$ is given
by (\ref{distr-x-basins}).
The $w_{il}$ occurring in this expression still have to be found.

\section{Solving the equations}
\label{solving}

We will now try and solve the problem of finding the weights $w_{il}$ of a
recurrent neural network, in the approximation dictated by equation
(\ref{averaging}) combined with the particular probability
distribution (\ref{prob-basins})--(\ref{prob-i-basins}), and we hope, thereby, to have obtained a useful solution for
the problem that we actually want to solve, i.e., the equations
(\ref{noise-basins}) or, equivalently, (\ref{noise-alt}) for given collections
$\Omega^{\mu}$.
The question to what extend we will have achieved this goal will be
answered in section~\ref{num-results}, where we perform a numerical
analysis.

The analytical approach to the problem to solve (\ref{averaging}), an
equation for the weights of a many neuron recurrent network is an adapted
version of the way in which Wiegerinck and Coolen calculated the weights
for a large perceptron \cite{wiegerinck}.

\subsection{Implicit equations for the weights}
\label{right}

By substituting (\ref{prob-basins})--(\ref{prob-i-basins}) into equation (\ref{averaging}), we can
obtain explicit expressions for both its left and its right side, and,
from these, solve for the weights $w_{il}$.
Using (\ref{average-x}), we immediately obtain for the right-hand side of (\ref{averaging})
\be
\label{right-final}
\sum_{\mu=1}^p  \sum_{\x \in \Omega^{\mu}}
p^{\mu}(\x) x_j \xi_i^{\mu} = \sum_{\mu} \xi_i^{\mu} \bar{x}_j^{\mu} 
\ee
where $ \bar{x}_j^{\mu}$ is given by (\ref{distr-x-basins}).
We turn now to the left-hand side of equation (\ref{averaging}), the
handling of which is more complicated and will be largely done in the
appendix.

We note that if $w_{ij}=w_{ij}(\theta_i,\xi_i^{\mu})$ is a solution of equation
(\ref{fixed}) or (\ref{noise-basins}), then also $\hat{w}_{ij}(\hat{\theta}_i,
\xi_i^{\mu}):= a_i w_{ij}(a_i^{-1}\hat{\theta_i},\xi_i^{\mu})$ is a
solution of equation (\ref{fixed}) or (\ref{noise-basins}), if $\theta_i$ is
replaced by $\hat{\theta}_i=a_i \theta_i$, where $a_i$ is an arbitrary real
constant.
Using this freedom of gauge with $a_i=(\sum_{m=1}^N w_{im}^2)^{1/2}$,
we can adjust the order of magnitude of the weights and the thresholds
\be
\label{norm}
\hat{w}_{ij}=\frac{w_{ij}}{\sqrt{\sum_{m=1}^N w_{im}^2}}  \qquad
\hat{\theta}_i=\frac{\theta_i}{\sqrt{\sum_{m=1}^N w_{im}^2}} 
\ee
which has a consequence that, if $w_{ij}$ and $\theta_i$ are of the
order $N^y$ ($y$ an arbitrary real number), the hatted
quantities are small, namely of the order $1/ \sqrt{N}$.
Note that 
\be
\label{sum=1}
\sum_{m=1}^N \hat{w}_{im}^2=1 \, .
\ee
The equations (\ref{norm}) and (\ref{sum=1}) enable us to switch, at
any moment, from hatted to unhatted quantities. The hatted quantities
are useful in view of the property (\ref{norm}), a property that is
used in the appendix.
One has, trivially,
\be
\label{theta-fixed}
\Theta_{\rm H}( \sum_{l=1}^N w_{il} x_l-\theta_i)=\Theta_{\rm H}( \sum_{l=1}^N
\hat{w}_{il} x_l-\hat{\theta}_i) \, .
\ee
The further evaluation of the left-hand side of (\ref{averaging}) in
terms of the $\hat{w}_{ij}$ is rather complicated and is given in the
appendix.
Combining the right-hand side, eq. (\ref{right-final}), and the left-hand
side, eq. (\ref{left-final}), we find an implicit equation for the $\hat{w}_{ij}$
\be
\label{w-3}
\hat{w}_{ij}=N^{-1} \sum_{\mu=1}^p E_i^{\mu}(b)  (2\xi_i^{\mu}-1)
\bar{x}_j^{\mu} \qquad (i=1,\ldots,N; j \in V_i)   
\ee
where the $E_i^{\mu}$ given by
\be
\label{e-basins}
E_i^{\mu}(b)=N \frac{(\gammabd)^{-1} \exp{(-(\gammabd)^2/ 2 \sigma )}}{\sum_{\mu} \exp{(-(\gammabd)^2/ 2 \sigma )}} 
\ee
are positive quantities.
In the latter equations we abbreviated $\sigma=b(1-b)$ and introduced
$\hat{\bar{\gamma}}_i^{\mu}(b,\w_i)$, quantities like the
$\bar{\gamma}_i^{\mu}$, equation (\ref{gamma-basins}), of which the
precise definition is given in the appendix by (\ref{gammabd}). 
With (\ref{w-3})--(\ref{e-basins}) we have obtained an expression for the
weights $\hat{w}_{ij}$ in terms of the $\gammabd$, which,
in turn, is a given function of the weights $\hat{w}_{ij}$, the
thresholds $\hat{\theta}_i$ and the patterns $\XI^{\mu}$.
In other words, the equations (\ref{w-3})--(\ref{e-basins}) are implicit
expressions for the weights only.

We could find explicit expressions for the weights by expanding
the $\hat{\bar{\gamma}}_i^{\mu}(b,\w_i)$ as a power series in the basin
parameter $b$
\be
\label{gamma-exp}
\hat{\bar{\gamma}}_i^{\mu}(b,\w_i)=\hat{\bar{\gamma}}_i^{\mu 0}
+\hat{\bar{\gamma}}_i^{\mu 1} b^1 + \hat{\bar{\gamma}}_i^{\mu 2} b^2 + \ldots
\ee
Inserting this expansion into (\ref{w-3})--(\ref{e-basins}), using
(\ref{sigma}), and equating equal powers of the expansion variable
$b$, we may obtain explicit expressions for the expansion coefficients
$\hat{\bar{\gamma}}_i^{\mu k}$
($\mu=1,\ldots,p;i=1,\ldots,N;k=0,1,2,\ldots,\infty$) of the power
series in $b$, in terms of the physical quantities
$\XI^{\mu}$, $\hat{\theta}_i$ and $\hat{w}_{ij}$, where $j$ is restricted to the
set $V_i^c$.
We thus would find an analytical solution of eq.~(\ref{averaging}).
This scheme has been carried out by Wiegerinck and Coolen \cite{wiegerinck} for
the perceptron.
We do not pursue this path for the recurrent neural net considered
here, but we will use a pragmatic shortcut to arrive at an approximate explicit expression
instead.
This will be done on the basis of an alternative implicit expression
for the weights (\ref{w-3}), to be derived in the next section [see
eq.~(\ref{w-alt}) below]. 

\subsection{An alternative implicit expression for the weights}
\label{alt-impl}

Rewriting (\ref{gammabd}), we may derive an alternative expression for
  $E_i^{\mu}(b)$.
To that end we substitute (\ref{w-3}) into (\ref{gammabd}):
\be
\label{bepalen-e}
\sum_{\nu=1}^p \bar{C}_i^{\mu \nu} E_i^{\nu}(b) (2
\xi_i^{\nu}-1)=\Gamma_i^{\mu}(b)  
\ee
where $\bar{C}_i^{\mu \nu}$ is the symmetric $p\times p$ correlation
matrix given by 
\be
\label{c}
\bar{C}_i^{\mu \nu}(b):=N^{-1} \sum_{m \in V_i} \bar{x}_m^{\mu} \bar{x}_m^{\nu}  
\ee
with $\mu,\nu=1,\ldots,p$ and where
\be
\label{hoofd-gamma}
\Gamma_i^{\mu}(b):= [\gammabd- (\sum_{m \in V_i^c} \hat{w}_{im}
\bar{x}_m^{\mu} -\hat{\theta}_i)(2 \xi_i^{\mu}-1)](2 \xi_i^{\mu}-1) \, .
\ee
From (\ref{bepalen-e}) we get, by multiplying both sides by $(2
\xi_i^{\lambda}-1) \bar{C}_i^{\mu \lambda}$ and summing over $ \lambda=1,\ldots,p$,
\be
\label{e-alt}
E_i^{\lambda}(b)= \sum_{\mu=1}^p \Gamma_i^{\mu}(b)(\bar{C}_i^{-1}(b))^{\mu
  \lambda} (2 \xi_i^{\lambda}-1) 
\ee 
where $\bar{C}^{-1}$ is the inverse of the matrix $\bar{C}$.
With (\ref{e-alt})) we have obtained an alternative expression for the $E_i^{\mu}(b)$
  [see equation (\ref{e-basins})] in terms of the same
  quantities, namely $\hat{w}_{ij}$, $\xi_i^{\mu}$, $ \hat{\theta}_i$ and $b$.
Substitution of this alternative expression (\ref{e-alt}) into
  (\ref{w-3}) leads to an alternative expression for the
  $\hat{w}_{ij}$ with $j \in V_i$:
\be
\label{w-alt}
\fl
\hat{w}_{ij} = N^{-1} \sum_{\mu,\nu=1}^p [\gammabd -
  (\sum_{m \in V_i^c} \hat{w}_{im} \bar{x}_m^{\mu} -\hat{\theta}_i) (2 \xi_i^{\mu}-1)] (2 \xi_i^{\mu}-1) (\bar{C}_i^{-1}(b))^{\mu \nu} \bar{x}_j^{\nu}
\ee
In equation (\ref{gammabd}) we introduced the $\gammabd$ as functions of the weights $\hat{w}_{ij}$.
Here, we have found, conversely, the weights in terms of the
  $\gammabd$.
By inserting $\hat{w}_{ij}$ (\ref{w-alt}) into $\gammabd$, equation
  (\ref{gammabd}), and making use of the definition (\ref{c}) for
  $\bar{C}_i^{\mu \nu}(b)$ one arrives, indeed, at an identity.
In view of (\ref{norm}), equation (\ref{w-alt}) also holds true with
  all hats dropped. 

The $\bar{\gamma}$'s occurring in (\ref{w-alt}) are given by
\be
\label{gamma-alt}
\eqalign{\bar{\gamma}_i^{\mu}(b,\w_i)= & N \sum_{\nu=1}^p \frac{\bar{C}_i^{\mu \nu}(b)
(2\xi_i^{\nu}-1) (2\xi_i^{\mu}-1) \exp{(- (\bar{\gamma}_i^{\nu}(b,\w_i))^2/ 2
    \sigma)}}{ \bar{\gamma}_i^{\nu}(b,\w_i) \sum_{\lambda}
  \exp{(-(\bar{\gamma}_i^{\lambda}(b,\w_i))^2/ 2 \sigma )}}\\  & +
(\sum_{m \in V_i^c} w_{im} \bar{x}_m^{\mu} -\theta_i)(2 \xi_i^{\mu}-1)  } 
\ee
as follows from (\ref{e-basins}), (\ref{bepalen-e}) and (\ref{hoofd-gamma}).
The equations (\ref{w-alt}) with (\ref{gamma-alt}) are an implicit
expression for the weights.
Developing the $\bar{\gamma}$'s according to (\ref{gamma-exp}), we
might obtain an explicit expression for the weights (\ref{w-alt}),
just as in section~\ref{right}.

The weights $w_{ij}$ have been constructed as a solution of equation
(\ref{averaging}), an equation which is strongly related to equation
(\ref{noise-alt2}).
Hence, one may expect that, on the average,
the $\gamma_i^{\mu}$'s are positive, i.e., 
\be
\label{gamma-pos}
\bar{\gamma}_i^{\mu}(b,\w_i)>0 \, .
\ee

We come now to the shortcut referred to above.
Instead of determining the coefficients of the expansion
(\ref{gamma-exp}) for the $\hat{\bar{\gamma}}$'s, we truncate this expansion
after the first term.
Dropping the hats and writing 
\be
\label{gamma-choice}
\bar{\gamma}_i^{\mu 0}=\kappa
\ee 
for \emph{all} constant first terms in the expansions
(\ref{gamma-exp}), we obtain from (\ref{w-alt})
\be
\label{w-expl}
\fl
w_{ij} = \cases{ \eqalign{ N^{-1} \sum_{\mu,\nu=1}^p & [\kappa -
  (\sum_{m \in V_i^c} w_{im} \bar{x}_m^{\mu} -\theta_i) (2
  \xi_i^{\mu}-1)] (2 \xi_i^{\mu}-1) \\
 & \times (\bar{C}_i^{-1}(b))^{\mu \nu} \bar{x}_j^{\nu} } &
($j \in V_i$)\\
w_{ij} \ \ \ \ ({\rm prescribed}) & ($j \in V_i^c$) }
\ee
Note that with the choice $w_{ij}(t_0)=0$ for $j \in V_i$ and
$w_{ij}(t_0)=w_{ij}$ (prescribed) for $j \in V_i^c$ in our main
result, eqs. (\ref{main})--(\ref{v-main}), the latter equations reduce
to the equations (\ref{w-expl}).
We thus have almost found the main result.
The final form (\ref{main})--(\ref{v-main}) is derived in section~\ref{otherwork}, after a
numerical analysis of the particular case (\ref{w-expl}).
  
In view of (\ref{gamma-pos}), we will choose for $\kappa$, in
eq. (\ref{gamma-choice}), a certain positive number.
This approach, in which we replace the constants $\bar{\gamma}_i^{\mu 0}$ by
a number to be found by (numerical) trial and error, is a priori,
rather crude.
The usefulness of this way of handling will be the subject of the next
section.

\section{Numerical results: probing the basins}
\label{num-results}

In this section we will study the question regarding
the size of the basins of attraction induced by the collection of patterns
$\Omega^{\mu}(b)$. 
Stated differently, we will determine whether the solution
(\ref{w-expl}) for the weights gives suitable basins of
attraction.
More in particular, we will search for the optimal values
$\kappa$ and $b$ to be taken in (\ref{w-expl}).
This will be done by carrying out a numerical analysis.

Let us denote, more extensively, the $\gamma_i^{\mu}$ of equation (\ref{noise-alt}) by $\gamma_i(\x,\w_i(b),\xi_i^{\mu})$.
Equation (\ref{noise-basins}), with weights $w_{ij}(b)$ given by
equations (\ref{w-expl}), is satisfied if, for a certain
pattern $\x$, the $\gamma_i(\x,\w_i(b),\xi_i^{\mu})$ are positive for
all $i$.
Therefore, we proceed as follows.

We construct probes consisting of patterns $\x$ centered around the
typical patterns $\XI^{\mu}$, and test whether these $\x$'s are
recognized by the neural net, i.e., we determine the sign of the
$\gamma_i$'s for the patterns $\x$ of the probe.
As a probing set we take patterns which are distributed around the
typical patterns $\XI^{\mu}$ in the same way as before, namely as
given by formulae (\ref{prob-basins}) and (\ref{prob-i-basins}), but now with the
basin-parameter $b$ replaced by a parameter $\bar{b}$. 
The latter parameter is dubbed `probing-parameter'.
In general, the probing-parameter $\bar{b}$ used in the test will be
unequal to the basin-parameter $b$ used to calculate the weights
$w_{ij}(b)$.
If the probing parameter $\bar{b}$ vanishes, a probing collection
$\Omega^{\mu}(\bar{b}=0)$ consists of precisely one pattern, namely
$\XI^{\mu}$.

In our numerical study, we first picked a certain value for the
probing-parameter $\bar{b}$, thereafter took an $\x$ belonging to the
probing set $\Omega^{\mu}(\bar{b})$ defined by this $\bar{b}$, and
thereupon calculated the
$\gamma_i(\x,\w_i(b),\xi_i^{\mu})$, equation (\ref{noise-alt}).
We repeated this procedure (for fixed $\bar{b}$) many times, and then
calculated the fraction of $\x$'s of the probing set for which all
$\gamma_i(\x,\w_i(b),\xi_i^{\mu})$ were positive.

\begin{figure}[htb]
\centerline{\hbox{\epsffile{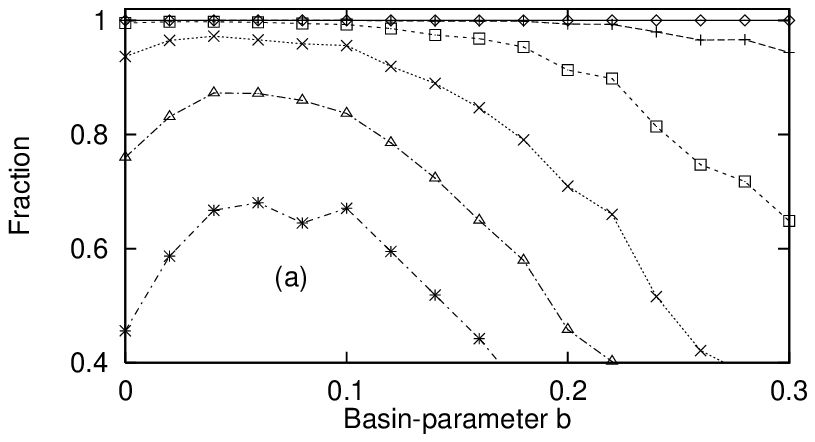} \epsffile{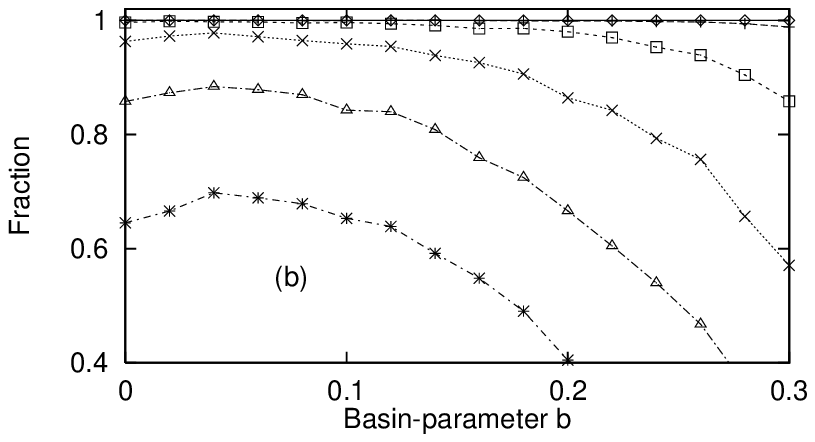}}}
\centerline{\hbox{\epsffile{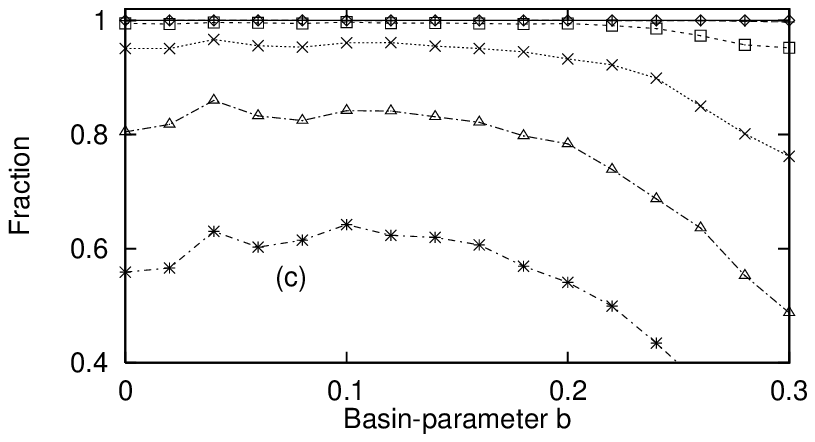} \epsffile{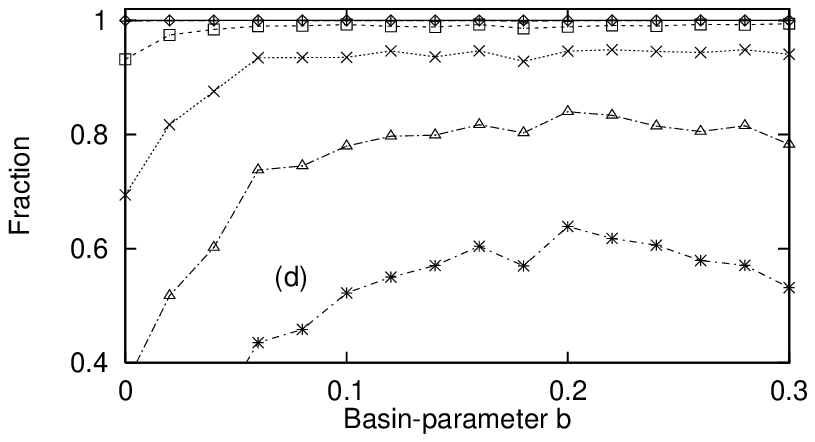}}}
\caption[0]{
\textbf{Probing of the basins for various values of the margin
  parameter}.\\ 
In the four graphs $a$, $b$, $c$ and $d$, the fraction of $x$'s with
  all $\gamma_i$ positive is depicted, vertically, for four values of the parameter $\kappa$
occurring in the final expression for the weights ($\kappa=1$,
$\kappa=2 N^{-1}$, $\kappa=N^{-1}$ and $\kappa=\frac{1}{2}N^{-1}$), as a
  function of the basin parameter $b$.
The six broken lines in each of the graphs correspond to different
values of the probing parameter $\bar{b}$ that characterize the sets
  $\Omega^{\mu}(\bar{b})$. 
From top to bottom, in each graph, we have plotted the fraction of
  $\x$'s with all $\gamma_i$ positive for values of $\bar{b}$ given by
  the six numbers $0$, $0.02$, $0.04$, $0.06$, $0.08$ and $0.1$, respectively.  
The number of neurons is $N=256$, the number of patterns $\XI$ equals
$p=32$.
The mean activity is $a=0.2$.
The dilution of the network is $d=0.2$.\\
In each of the four graphs, $a$, $b$, $c$ and $d$, that is, for four
  different values of the margin parameter $\kappa$, there is an
  interval of values of $b$ for which the fraction of $\gamma$'s
  equals one, for a range of values of the probing parameter
  $\bar{b}$.
Hence, for probes with $\bar{b}$ in the latter range, the net has
  values for the weights $w_{ij}(b)$ which are such that the net
  performs optimally.
\protect\label{fignoise1}}
\end{figure}

In figure~\ref{fignoise1}, we have depicted the relative number of
$\x$'s belonging to the basin (vertical axis) as a function of the
basin-parameter $b$ (horizontal axis).
The graphs $a$, $b$, $c$ and $d$ in figure~\ref{fignoise1} correspond to
four values of the margin parameter $\kappa$: $\kappa=1$,
$\kappa=2 N^{-1}$, $\kappa=N^{-1}$ and $\kappa=\frac{1}{2}N^{-1}$.
All patterns $\XI^{\mu}$ are supposed to have the property that an
arbitrary chosen $\xi_i^{\mu}$ has probability $a$ to be equal to $1$.
This probability $a$ is referred to as the mean activity.
Note that for random patterns the mean activity is given by $a=0.5$.
Experimentally, however, the mean activity is found to be smaller
\cite{abelles}.
In all graphs we have chosen vanishing prescribed weights, $w_{ij}=0$, $j
\in V_i^c$, and $\theta_i=N^{-1}$ for all
$i=1,\ldots,N$.
That is, we considered diluted networks.
More specifically, we took, randomly, $20$ percent of the weights to belong to the
set $V_i^c$, which corresponds to a dilution $d=0.2$.

Each of the broken lines in the graphs \ref{fignoise1}$a$--
\ref{fignoise1}$d$ corresponds to a different value of the probing parameter $\bar{b}$.
Going from top to bottom in the four graphs of figure~\ref{fignoise1},
we cross curves with a larger and larger probing parameter $\bar{b}$.
For the smallest possible value of the probing-parameter $\bar{b}$,
namely $\bar{b}=0$, the probing set reduces to a typical pattern $\XI^{\mu}$.
It follows from figure~\ref{fignoise1} (see the
upper lines, little diamonds) that the fraction of $\x$'s belonging to
a basin equals $1$ for a large range of the basin-parameter $b$.
As is to be expected, a typical pattern $\XI^{\mu}$ indeed is a fixed
point for all values of $b$ (up to some upper limit which is larger
than $0.3$).

For values of the probing-parameter $\bar{b}$ close to zero,
$\bar{b}=0.02$ say, the fraction of $\x$'s belonging to a basin equals
one for a large range of the basin parameter $b$ (see the second curves
from above, indicated by little plus signs).
As long as the probing-set is smaller than the set of patterns which
belong to the basin of attraction, the fraction remains one. 
In case this fraction is less than one, the probing-set is larger than
the set of patterns which form the basins of attraction.
Hence, the probing-parameter $\bar{b}$ can be viewed upon as a measure for the
size of the basin of attraction.

To illustrate these latter statements we take as an example figure
$1d$.
The lines $\bar{b}=0$ and $\bar{b}=0.02$ coincide: they are the
straight horizontal line with fraction one.
For $\bar{b}=0.04$, corresponding to a fraction given by the curve
with little squares, the fraction rises to one as a function of $b$.
This implies that the size of the basins grows as a function of $b$.
For larger values of $\bar{b}$, given by the curves with crosses,
triangles and asterisks, the fraction also rises as a function of $b$,
up to some value of $b$, but never equals one.
So in these cases, the number of elements of the probing sets always
clearly is larger than the number of elements belonging to the basins.

Now, we come to the effect of $\kappa$ on the performance of the
network.
Comparing figures $1a$ and $1d$, and looking where the fraction equals
one, we discover that for large $\kappa$, $b$ should be small, and
vice versa.

\begin{figure}[htb]
\centerline{\hbox{\epsffile{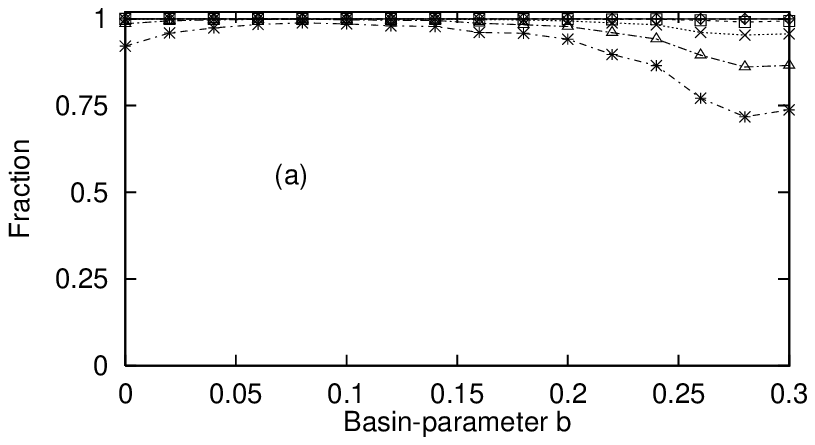} \epsffile{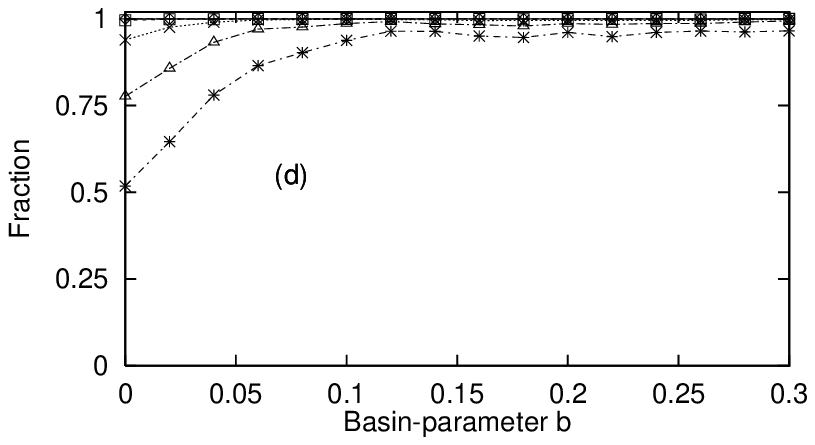}}}
\centerline{\hbox{\epsffile{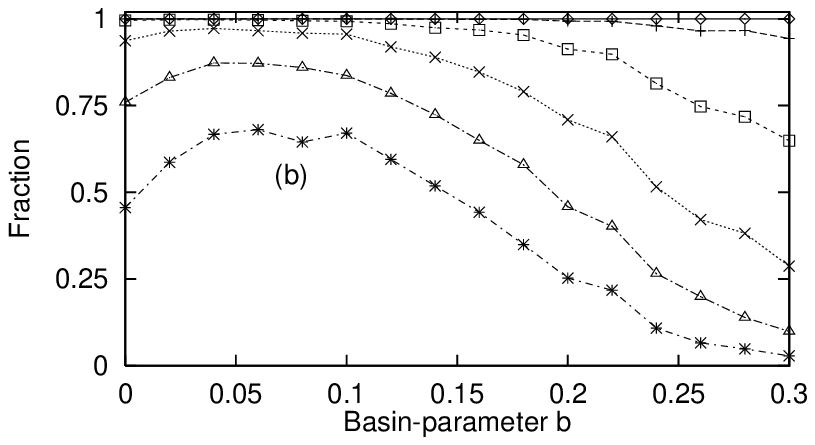} \epsffile{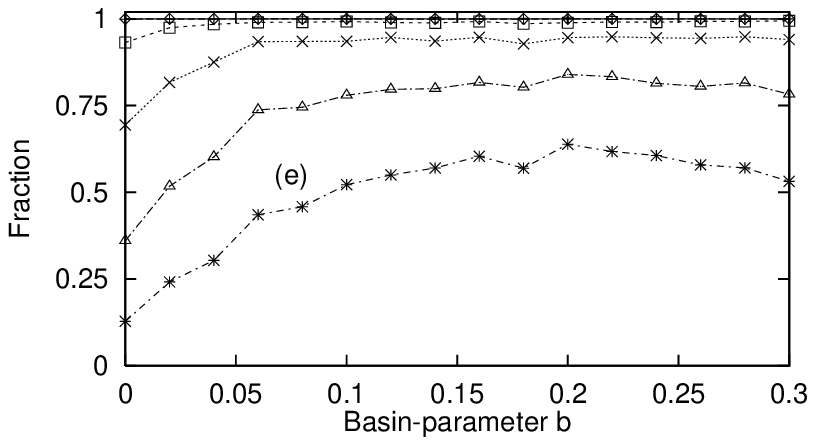}}}
\centerline{\hbox{\epsffile{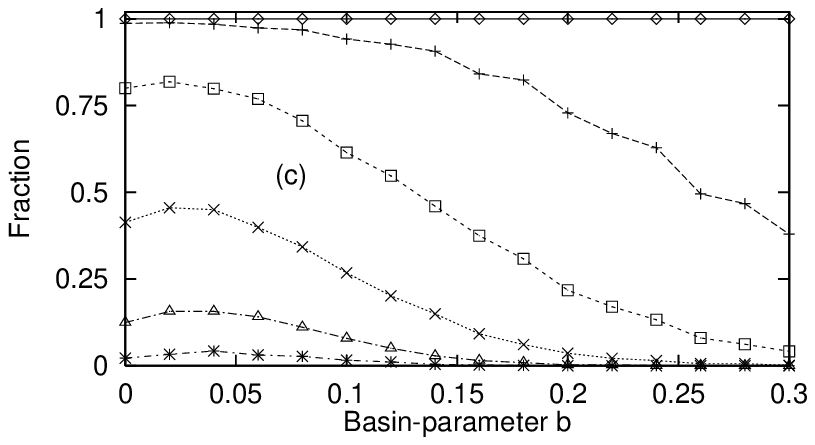} \epsffile{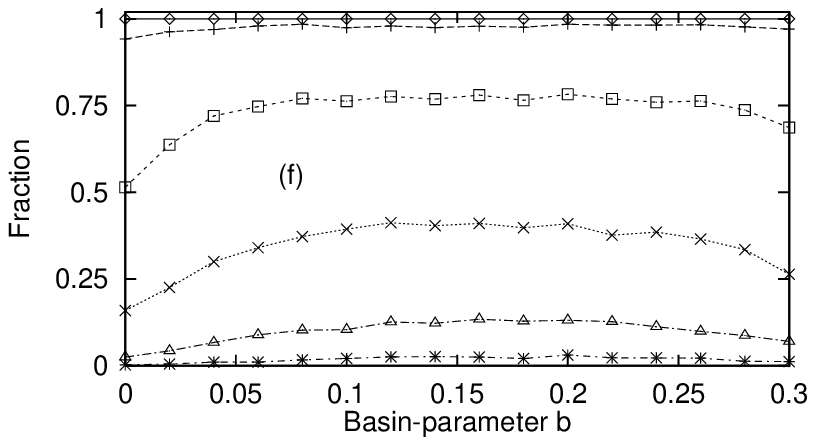}}}
\caption[1]{
\textbf{Probing of the basins for various numbers of patterns}.\\
The fraction of $\x$'s with all $\gamma_i$ positive is depicted, vertically, for
three different values $p$ of the number of stored patterns, $p=16$ (top),
$p=32$ and $p=64$ (bottom), as a function of the basin parameter $b$.
In the left-column the margin parameter is chosen large compared to
  the threshold, $\kappa=1$, whereas in the right-column $\kappa$ is
  taken of the order of the threshold, $\kappa=\frac{1}{2}N^{-1}$. 
The six broken lines in each of the graphs correspond to different values of the probing-parameter $\bar{b}$.
From top to bottom in each graph we have plotted the fraction of
  $\x$'s with all $\gamma_i$ positive for values of $\bar{b}$ given by
  $0$, $0.02$, $0.04$, $0.06$, $0.08$ and $0.1$, respectively.  
The number of neurons is $N=256$, the mean activity is $a=0.2$.
The dilution of the network is $d=0.2$.\\
It is seen that for $b \neq 0$, the fraction rises, up to some value
of $b$.
Hence, for large $\kappa$ (left column) and small $\kappa$ (right
column), the net performs better for $b \neq 0$, for different values
of the number of patterns $p$.
\protect\label{fignoise2}}
\end{figure}

In figure~\ref{fignoise2}, we study for a large value $\kappa=1$ and a
small value $\kappa=\frac{1}{2}N^{-1}$ of the margin parameter what happens
when the number of patterns varies from $16$ via $32$ to $64$.
As before we have taken vanishing prescribed weights, $w_{ij}=0$, $j
\in V_i^c$, $\theta_i=N^{-1}$ for all
$i=1,\ldots,N$, and dilution $d=0.2$.
We find for $\kappa=1$ as well as $\kappa=\frac{1}{2}N^{-1}$ that when
the number of patterns increases, the size of the basins decreases.
But, since the curves have a hump, a value
for the basin parameter $b$ unequal zero yields a network that
recognizes a larger part of the probing sets $\Omega^{\mu}(\bar{b})$.

The final observation relating to figures~\ref{fignoise1} and
\ref{fignoise2} reads that, in general, a network with weights
$w_{ij}(b \neq 0)$ possesses larger basins of attraction than a
network with weights $w_{ij}(b=0)$.

\section{Relation to earlier work}
\label{otherwork}

The above mathematical study has been performed for adaptable weights,
$w_{ij}$, $j \in V_i$, to be determined by the equations (\ref{averaging}),
and prescribed weights, $w_{ij}$, $j \in V_i^c$.
Let us turn to the situation of a neural network that adapts its
weights, in the course of time, according to some learning rule.
In such a network, all weights start, at $t=t_0$ say, with some
initial value $w_{ij}(t_0)$.
The weights $w_{ij}$, with $j \in V_i^c$, keep their weights throughout
the learning process, while the weights $w_{ij}$, with $j \in V_i$, change in
the course of time.
Now, we ask the question whether we can find $\tilde{w}_{ij}$ which
are such that $\tilde{w}_{ij}(t)$ has prescribed values $w_{ij}(t_0)$,
for all $i$ and $j$, at $t=t_0$, whereas
$\bar{\gamma}_i^{\mu}(b,\tilde{\w}_i(t))$ has a large probability of
being positive.
One way to obtain these $\tilde{w}_{ij}$ is via the $w_{ij}$'s that
are given by the unhatted counterpart of equation (\ref{w-alt}).
In fact, they are given by
\be
\label{w-learn}
\tilde{w}_{ij}(t) =  \cases{w_{ij}(t_0)+ v_{ij}(t) & ($j \in V_i$)\\ 
w_{ij}(t_0) & ($j \in V_i^c$) }
\ee
where
\be
\label{basin-v}
v_{ij}(t)=w_{ij}(t) - N^{-1} \sum_{\mu,\nu=1}^p \sum_{m \in V_i} w_{im}(t_0)
\bar{x}_m^{\mu} \bar{C}_i^{-1}(b))^{\mu \nu} \bar{x}_j^{\nu} 
\ee
in which we have denoted the (unhatted counterparts of) $w_{ij}$ of equation (\ref{w-alt}) as
$w_{ij}(t)$.
An alternative way to write equation (\ref{basin-v}) is given by
\be
\label{v-alt}
v_{ij}(t)= N^{-1} \sum_{\mu,\nu=1}^p [\bar{\gamma}_i^{\mu}(b,\w_i(t))-
  \bar{\gamma}_i^{\mu}(b,\w_i(t_0))](2 \xi_i^{\mu}-1)  (\bar{C}_i^{-1}(b))^{\mu
  \nu} \bar{x}_j^{\nu} \, .
\ee
The weights $\tilde{w}_{ij}$, equation (\ref{w-learn}), have been constructed in such a way that
\be
\label{gam=gam}
\bar{\gamma}_i^{\mu}(b,\tilde{\w}_i(t))=\bar{\gamma}_i^{\mu}(b,\w_i(t)) \, . 
\ee
The latter equation can be verified easily.
In fact, inserting (\ref{w-learn}) with (\ref{v-alt}) into
(\ref{gamma-basins}) gives 
\be
\label{to_gam=gam}
\eqalign{ \bar{\gamma}_i^{\mu}(b,\tilde{\w}_i(t)) =
  \bar{\gamma}_i^{\mu}(b,\w_i(t_0)) + \sum_{\nu,\lambda=1}^p &
  [\bar{\gamma}_i^{\nu}(b,\w_i(t)) - \bar{\gamma}_i^{\nu}(b,\w_i(t_0))](2
  \xi_i^{\nu}-1) \\
& \times(2 \xi_i^{\mu}-1)(\bar{C}_i^{-1}(b))^{\lambda \nu} \bar{C}_i^{\lambda \mu}(b) } 
\ee
where we used the definitions (\ref{gamma-basins}) and (\ref{c}).
Since $ \bar{C}_i^{\lambda \mu}(b)$ is symmetric, the product of the
matrices $\bar{C}$ gives a Kronecker delta, which in turn yields
  (\ref{gam=gam}).
The property (\ref{gam=gam}) guarantees that when the
$\bar{\gamma}_i^{\mu}(b,\w_i(t))$ are positive, the
$\bar{\gamma}_i^{\mu}(b,\tilde{\w}_i(t))$ are also positive.

Using the same shortcut as above, equation (\ref{gamma-choice}), we
obtain 
\be
\label{v-final}
v_{ij}(t)= N^{-1} \sum_{\mu,\nu=1}^p [\kappa-
  \bar{\gamma}_i^{\mu}(b,\w_i(t_0))](2 \xi_i^{\mu}-1)  (\bar{C}_i^{-1}(b))^{\mu
  \nu} \bar{x}_j^{\nu} 
\ee
with $\bar{x}_j^{\nu}$ given by (\ref{distr-x-basins}).
The equations (\ref{w-learn}) with (\ref{v-final}) are equivalent to
  the main result (\ref{main})--(\ref{v-main}) mentioned in the introduction.
Putting in this expression the basin parameter equal to zero ($b=0$),
  we recover the expression obtained after a learning process in a
  preceding article \cite{heerema}.
This suggest that (\ref{w-learn}) with (\ref{v-final}) is the
  generalization of the weights in a process of learning with noisy
  patterns.
Hence, we may state that a network performs optimally when trained with
noise ($b \neq 0$), or, stated differently (and less precise): a
neural network performs best in an environment identical to the
training environment.
This is what Wong and Sherrington refer to as the `principle of
adaptation' \cite {ws}.
In a next article, we will extensively come back to this question, in a
  biological context \cite{heerema3}.
The final result will turn out to be that the expression
  (\ref{w-learn}) with (\ref{v-final}) is, apart from a detail, indeed
  the generalization of learning with noisy patterns.

\section{Conclusion}

Although we studied a neural network, we did not consider learning and
learning rules.
We simply asked the question: what values must one take for the
weights of a neural network in order that it performs optimally, i.e.,
that it can retrieve the largest sets of perturbed patterns.
We were able to reformulate this problem in a mathematically exact
way, and to obtain a solution that, by its construction, had a certain
plausibility of being a suitable one.
Finally, we performed a numerical test, which confirmed the usefulness
of our approach.
The weights $w_{ij}(b)$ obtained in this article on
the basis of perturbed data ($b \neq 0$), yield a network with larger
basins than would have been obtained in case of non-perturbed data
($b=0$).
In a subsequent article we will propose a biological learning rule,
which is such that, apart from a minor detail, the synapses strive at
the values given by the main result of this article, equations (\ref{main})--(\ref{v-main}).
In other words, nature might realize almost totally what mathematics suggests.

\appendix

\section{Derivation of implicit equations for the weights}
\label{lefthandside}

In this appendix we will evaluate the left-hand side of equation
(\ref{averaging}).
Then combining this with the result of section~\ref{right} for the right-hand
side will lead to implicit equations for $\hat{w}_{ij}$.

Inserting (\ref{theta-fixed}) into the left-hand side of
(\ref{averaging}), multiplying by a delta function containing a
variable $z$ and integrating over $z$, we get the
equivalent expression
\be
\label{left1}
\eqalign{ & \sum_{\mu} \sum_{x_1=0,1} \ldots \sum_{x_N=0,1} p^{\mu}(\x)  \int \! dz \, x_j \Theta_{\rm H}( \hat{w}_{ij} x_j-\hat{\theta}_i +z) \delta[z-
\sum_{l \neq j} \hat{w}_{il} x_l] \\
&= \sum_{\mu} \int \! dz \, \sum_{x_j} p_j^{\mu}(x_j) x_j
\Theta_{\rm H}( \hat{w}_{ij} x_j-\hat{\theta}_i +z) P_{ij}^{\mu}(z) }
\ee
where we used (\ref{prob-basins}) and where we abbreviated
\be
\label{p-z}
P_{ij}^{\mu}(z)= \sum_{x_1} \ldots \sum_{x_{j-1}} \sum_{x_{j+1}}
\ldots \sum_{x_N} \prod_{m \neq j} p_m^{\mu}(x_m) \delta[z-
\sum_{l \neq j} \hat{w}_{il} x_l]  \qquad (j \in V_i) \, .
\ee
The summation over $x_j$ in (\ref{left1}) yields
\be
\label{left2}
\sum_{x_j} p_j^{\mu}(x_j) x_j \Theta_{\rm H}( \hat{w}_{ij}
x_j-\hat{\theta}_i +z)= \bar{x}_j^{\mu} \Theta_{\rm H}( \hat{w}_{ij} -\hat{\theta}_i+z)
\ee
as follows by inserting (\ref{prob-i-basins}).
The factor $P_{ij}^{\mu}(z)$ can be rewritten in the following way.

Using a well-known representation of the delta-function we first
obtain
\be
\label{p-z1}
P_{ij}^{\mu}(z)= \frac{1}{2\pi} \int_{- \infty}^{\infty} \! dk \, {\rm
  e}^{ikz} \prod_{m \neq j}
\sum_{x_m} p_m^{\mu}(x_m) {\rm e}^{-ik \hat{w}_{im} x_m} \, .
\ee
One has
\be
\label{p-z2}
\sum_{x_m} p_m^{\mu}(x_m) {\rm e}^{-ik \hat{w}_{im} x_m}= (1-b) {\rm e}^{-ik \hat{w}_{im}
  \xi_m^{\mu}} + b {\rm e}^{-ik \hat{w}_{im} (1-\xi_m^{\mu})} 
\ee
where we used (\ref{prob-i-basins}).
Inserting (\ref{p-z2}) into (\ref{p-z1}) we may write
\be
\label{p-z3}
P_{ij}^{\mu}(z)= \frac{1}{2\pi} \int \! dk \, \exp{ \{ ikz + \sum_{m \neq j}
\ln{[ (1-b) {\rm e}^{-ik \hat{w}_{im} \xi_m^{\mu}} + b {\rm e}^{-ik \hat{w}_{im}
  (1-\xi_m^{\mu})} ]} \} } 
\ee
where we used $y=\exp{\{ \ln{y} \}}$.
We can now expand the two exponentials occurring in the argument of
the logarithm.
This leads to a term of the form $\ln{(1+y)}$.
Thereupon, we can expand this term as $y-\frac{1}{2}y^2+\ldots$, since $y$ is of
the order of $\hat{w}_{ij}$, and $\hat{w}_{ij}$ is of the order
$N^{-1/2}$, as noted above [see eqs. (\ref{norm}) and following text].
Thus we obtain
\be
\label{ln}
\ln{[ (1-b) {\rm e}^{-ik \hat{w}_{im} \xi_m^{\mu}} + b {\rm e}^{-ik \hat{w}_{im}
    (1-\xi_m^{\mu})} ]} = -ik\hat{w}_{im} \bar{x}_m^{\mu} -\frac{1}{2} b(1-b) k^2 \hat{w}_{im}^2 + \ldots
\ee
Inserting (\ref{ln}) into (\ref{p-z3}) we may write
\be
\label{p-z4}
P_{ij}^{\mu}(z)= \frac{1}{2\pi} \exp{\{ -(z-z_0)^2 /2 \sigma \}}
 \int_{- \infty}^{\infty} \! dk \, \exp{ \{
 -\frac{\sigma}{2}(k-i(z-z_0)/ \sigma )^2 \}} + \ldots 
\ee
where we abbreviated
\be
\label{sigma-z0}
\sigma:=b(1-b) \sum_{m \neq j} \hat{w}_{im}^2 \qquad z_0:=\sum_{m \neq
 j} \hat{w}_{im} \bar{x}_m^{\mu} \, .
\ee
Using the fact that $\hat{w}_{ij}$ is of the order $1/\sqrt{N}$ we may
 write
\be
\label{sigma}
\sigma=b(1-b)
\ee
a relation we will use later.
After evaluating the integral (\ref{p-z4}), we obtain
\be
\label{p-final}
P_{ij}^{\mu}(z)= (2 \pi \sigma)^{-\frac{1}{2}} \exp{\{
 -(z-z_0)^2/2 \sigma \}} + \ldots  \qquad (i=1,\ldots,N; j \in V_i) 
\ee
with $\mu=1,\ldots,p$.
Substituting (\ref{left2}) and (\ref{p-final}) into the right-hand
 side of (\ref{left1}) we obtain for the left-hand side of (\ref{averaging}) 
\be
\label{left3}
(2 \pi \sigma)^{-\frac{1}{2}} \sum_{\mu} \int \! dz \, \bar{x}_j^{\mu} \Theta_{\rm H}( \hat{w}_{ij}-\hat{\theta}_i +z) \exp{\{ -(z-z_0)^2/2 \sigma \}} \, .
\ee
The integral occurring in (\ref{left3}) can be rewritten
\be
\label{i-def}
I_{ij}^{\mu}:= (2 \pi \sigma)^{-\frac{1}{2}} \int \! dz \, \Theta_{\rm H}(
 \hat{w}_{ij}-\hat{\theta}_i +z) \exp{\{ -(z-z_0)^2/2 \sigma \}} \, .
\ee
Changing the integration variable $z$ according to $y=(z-z_0)/\sqrt{2
 \sigma}$, we find
\bea
\label{i2}
\fl 
I_{ij}^{\mu} &=& \pi^{-\frac{1}{2}} \int_{-\infty}^{\infty} \! dy
\, \Theta_{\rm H}( \hat{w}_{ij}-\hat{\theta}_i +z_0 +\sqrt{2 \sigma}y)
{\rm e}^{-y^2} \nn\\ 
&=&  \pi^{-\frac{1}{2}} \int_0^{\infty} \! dy \, {\rm e}^{-y^2} + (4 \pi)^{-\frac{1}{2}} \int_0^{\infty} \! dy \, [{\rm
  sgn}(\hat{w}_{ij}-\hat{\theta}_i +z_0 -\sqrt{2 \sigma}y) \nn\\
&& + {\rm sgn}(\hat{w}_{ij}-\hat{\theta}_i+z_0  +\sqrt{2 \sigma}y)] {\rm
  e}^{-y^2} \, .
\eea
The integral over the first term is a Gaussian integral, the second
term can be expressed in an error function.
We obtain
\be
\label{i3}
I_{ij}^{\mu}=\frac{1}{2}+ \frac{1}{2} {\rm erf}([\gammabd
(2\xi_i^{\mu}-1)+\epsilon_{ij}^{\mu}]/ \sqrt{2 \sigma}) \qquad
(i=1,\ldots,N; j \in V_i)  
\ee
where $\mu=1,\ldots,p$ and where the error function is defined according to
\be
{\rm erf}(x):= \frac{2}{\sqrt{\pi}} \int_0^{x} \! dy \, {\rm e}^{-y^2}
\, .
\ee
In analogy to (\ref{gamma-basins}) we defined
\be
\label{gammabd}
\gammabd= (\sum_{l=1}^N \hat{w}_{il} \bar{x}_l^{\mu} -\hat{\theta}_i)(2 \xi_i^{\mu}-1) \, .
\ee
Furthermore, we abbreviated
\be
\label{epsilon}
\epsilon_{ij}^{\mu}= -\hat{w}_{ij} \bar{x}_j^{\mu}+ \hat{w}_{ij} \, .
\ee
Note that, apart from a $\XI^{\mu}$ dependent factor, the quantity $\epsilon_{ij}^{\mu}$ equals the
weight $\hat{w}_{ij}$.
In view of (\ref{norm}), $\epsilon_{ij}^{\mu}/ \sqrt{2 \sigma}$ is
small.
The error function in (\ref{i3}) can be split into two contributions.
For small $\epsilon$ we have
\be
\int_\gamma^{\gamma+\epsilon} \! dy \, {\rm e}^{-y^2}= \epsilon {\rm e}^{-\gamma^2} + \ldots
\ee
which allows us to write for (\ref{i3})
\be
\label{i4}
\fl
I_{ij}^{\mu}=  \frac{1}{2}+ \frac{1}{2} {\rm erf}(\gammabd(2 \xi_i^{\mu}-1)/ \sqrt{2 \sigma}) +
\frac{\epsilon_{ij}^{\mu}}{\sqrt{2 \pi \sigma}} \exp{( -(\gammabd)^2/
  2 \sigma )} + \ldots
\ee
Using (\ref{left3}) and (\ref{i4}) with (\ref{gammabd}), the final expression for the
left-hand side of (\ref{averaging}) can be obtained
\be
\label{left-final}
\fl
\frac{1}{2} \sum_{\mu} \bar{x}_j^{\mu} [1+ {\rm
  erf}(\gammabd(2 \xi_i^{\mu}-1)/ \sqrt{2 \sigma})] + \frac{b(1-b)\hat{w}_{ij}}{\sqrt{2 \pi \sigma }} \sum_{\mu} \exp{(-(\gammabd)^2/ 2\sigma )}
\ee
Combining the right- and left-hand sides of equation
(\ref{averaging}), as given by (\ref{right-final}) and
  (\ref{left-final}), respectively, we get an equation from which the
  weights $\hat{w}_{ij}$ follow immediately
\be
\label{w-1}
\hat{w}_{ij}= \frac{ \sqrt{2 \pi \sigma} \sum_{\mu} \bar{x}_j^{\mu} [(2\xi_i^{\mu}-1) - {\rm
  erf}(\gammabd(2 \xi_i^{\mu}-1)/ \sqrt{2\sigma})] } {2b(1-b)
  \sum_{\mu} \exp{(-(\gammabd)^2/ 2 \sigma )}} \, .
\ee
With the properties
\be
{\rm erf}(\gammabd (2\xi_i^{\mu}-1)/ \sqrt{2 \sigma})=(2\xi_i^{\mu}-1) {\rm erf}(\gammabd/ \sqrt{2 \sigma})
\ee
and
\be
{\rm erf}(y)=1- \frac{1}{y \sqrt{\pi}} {\rm e}^{- y^2} + \ldots
\ee
we can rewrite (\ref{w-1}),
\be
\label{w-2}
\eqalign{\hat{w}_{ij} = \frac{\sqrt{2 \pi \sigma}}{2b(1-b)} \sum_{\mu}
  & \bar{x}_j^{\mu} (2\xi_i^{\mu}-1) [\sqrt{\pi /
  2\sigma} \gammabd ]^{-1} \\
& \times \exp{(-(\gammabd)^2/ 2 \sigma )}/  \sum_{\mu}
  \exp{(-(\gammabd)^2/ 2 \sigma )} }
\ee
or, equivalently, (\ref{w-3}) with (\ref{e-basins}), the final results
  to be obtained in this appendix.

\end{document}